\documentclass[a4paper,11pt]{article}
\usepackage{jinstpub} 
\usepackage{lineno}
\usepackage{amsmath,amsthm,bm}
\usepackage{mathtools}
\usepackage{enumitem}
\usepackage{subcaption}
\usepackage{upgreek}

\newcommand{\rtrue}{\mathbf r_{\mathrm{true}}}
\newcommand{\rhat}{\hat{\mathbf r}}

\title{\boldmath Reconstruction Bias in Timepix4 Subpixel Centroiding for Electron Imaging}

\author[1]{N. Dimova\note{Corresponding author.}}
\author{R. Plackett}
\author{D. Bortoletto}
\affiliation{University of Oxford,\\
Department of Physics, Parks Road, Oxford OX1 3PU, UK}

\emailAdd{nina.dimova@physics.ox.ac.uk}

\abstract{
Subpixel centroiding is widely used to improve the spatial resolution of hybrid pixel detectors for electron imaging by estimating the true interaction position within an entry pixel. Existing centroiding strategies are typically optimised using localisation accuracy. However, observations show that improved localisation does not necessarily translate into improved modulation transfer function (MTF) or reliable virtual-pixel rebinning. This work proposes a unified interpretation of these observations based on the ambiguity of the centroid reconstruction problem. We show that observable-based centroid estimators solve an intrinsically non-unique inverse problem using deterministic reconstruction rules, which can introduce entry-phase-dependent reconstruction bias that governs the spatial distribution of reconstructed subpixel coordinates. This framework explains why localisation accuracy alone is insufficient to predict imaging performance and identifies approximate intra-pixel translational symmetry as a prerequisite for faithful virtual-pixel imaging. The proposed interpretation is evaluated using simulated 200~keV and 300~keV electron data together with measured 200~keV data acquired with a Timepix4 detector. Comparisons of charge-weighted, timing-based, and morphology-dependent centroiding strategies demonstrate that estimators with similar localisation performance can exhibit markedly different MTFs and rebinned flat-field behaviour. The results suggest that future centroid optimisation should consider subpixel phase bias and rebinnability alongside localisation accuracy.
}


\begin{document}
\maketitle
\flushbottom

\section{Introduction}

Hybrid pixel detectors are increasingly used for direct electron imaging in transmission electron microscopy (TEM)~\cite{FARUQI2018180}. In this imaging mode, an incident electron deposits charge across multiple neighbouring pixels, forming a cluster whose spatial and temporal structure depends on the electron energy, scattering history, charge transport within the sensor, and detector electronics response. Subpixel centroiding seeks to reconstruct the electron entry point from these cluster observables, potentially allowing the reconstructed image to be rebinned onto a finer virtual pixel grid and thereby improve the effective imaging resolution beyond the native detector pitch. Existing centroiding methods are typically optimised using localisation metrics, such as the distance between the reconstructed and true interaction positions in simulation~\cite{VANSCHAYCK2020113091, Xie_2024}.

However, localisation accuracy alone does not fully characterise imaging performance. In practice, centroid estimators with smaller localisation residuals do not necessarily produce higher modulation transfer functions (MTFs), while some methods generate systematic artefacts when data is rebinned into virtual pixels~\cite{VANSCHAYCK2020113091, Khalil}. These observations indicate that detector imaging performance depends not only on localisation accuracy, but also on the intra-pixel spatial organisation of the reconstructed centroid coordinates. The severity of the resulting reconstruction bias depends on the information content of the measured cluster. In particular, detectors that record substantially larger clusters~\cite{Xie_2024} are expected to support reconstruction with less severe phase-dependent bias.

This work provides a unified interpretation of these observations. We show that observable-based centroid estimators solve an intrinsically non-unique inverse problem using deterministic reconstruction rules and so can introduce reconstruction bias dependent on the true electron entry phase within the pixel. This framework explains why localisation accuracy alone does not predict MTF, why certain centroiding strategies cannot be successfully rebinned despite improved localisation performance, and why approximate intra-pixel translational symmetry is a necessary condition for faithful virtual-pixel imaging.

The proposed framework is evaluated using simulated 200~keV and 300~keV
electron data generated with Allpix$^2$~\cite{Allpix}, together with measured
200~keV electron data acquired using a $300~\upmu$m silicon sensor bump-bonded
to a Timepix4 ASIC~\cite{TPX4}.  Charge-weighted, timing-based, and
morphology-dependent centroiding strategies are compared in terms of localisation accuracy, modulation transfer function, and rebinned flat-field response.

\section{Fundamental ambiguity of centroid reconstruction}
\label{sec:2}

\subsection{Subpixel phase and reconstructed density}

The true electron interaction position may be written as the sum of an integer pixel coordinate and a continuous subpixel phase,
\begin{align}
\rtrue = (x_0 + \phi_x,\; y_0 + \phi_y),
\qquad
\phi_x,\phi_y \in [-0.5,0.5),
\end{align}
where $(\phi_x,\phi_y)$ specifies the interaction position within the pixel cell. For a uniformly illuminated flat field, the true subpixel phase distribution is uniform,
\begin{align*}
p(\phi_x,\phi_y)=1, \qquad (\phi_x,\phi_y)\in[-0.5,0.5)^2.
\end{align*}

A measured event cluster is represented by the pixel coordinates together with the measured Time over Threshold (ToT), denoted by $q$, and Time of Arrival (ToA), denoted by $t$, recorded in each hit pixel. A centroid estimator maps these measured cluster observables,
\begin{align*}
Z=\{(x_i,y_i,q_i,t_i)\}_{i=1}^{N}
\end{align*}
to a reconstructed position,
\begin{equation}
    \rhat=f(Z).
    \label{E2.4}
\end{equation}

The reconstruction is characterised by the conditional response
\begin{align*}
p(\rhat\mid\rtrue),
\end{align*}
which describes the probability of reconstructing a centroid at $\rhat$ for a true interaction at $\rtrue$. The reconstructed subpixel phase density is therefore
\begin{align}
p(\hat{\phi}_x,\hat{\phi}_y) = \int_{[-0.5,0.5)^2} p(\hat{\phi}_x,\hat{\phi}_y\mid\phi_x,\phi_y)\, p(\phi_x,\phi_y)\, d\phi_x\,d\phi_y.
\end{align}
Since the true subpixel phase distribution $p(\phi_x,\phi_y)$ is uniform, any structure in the reconstructed phase density must originate from the centroid estimator itself. Preferred reconstructed subpixel phases therefore appear as periodic occupancy modulations after rebinning. Throughout this work, these systematic departures from a uniform reconstructed phase density are referred to as \emph{subpixel phase bias}.

\subsection{Why centroid reconstruction is fundamentally ambiguous}

The detector does not observe the microscopic interaction directly. Instead, the measured cluster observables $Z=\{(x_i,y_i,q_i,t_i)\}_{i=1}^{N}$ arise after stochastic scattering, charge transport, thresholding, and digitisation. Consequently, the detector response is described by the conditional distribution $p(Z\mid\phi_x,\phi_y)$, rather than by a unique mapping. Ambiguity arises from both stochastic detector physics, where a single true interaction phase can produce multiple possible detector responses, and irreversible information loss during measurement, where different microscopic interactions can produce statistically indistinguishable measured observables.

Because charge sharing, threshold crossing, and cluster topology all depend on the electron entry phase, the detector response $p(Z\mid\phi_x,\phi_y)$ is itself phase dependent. The centroid estimator therefore does not receive statistically equivalent inputs at every subpixel phase. Even if the estimator applies a smooth deterministic reconstruction rule, the distribution of its inputs remains phase dependent. The inverse reconstruction problem is therefore inherently probabilistic and does not admit a unique solution, even if the detector physics is known exactly.

\subsection{Phase bias as a consequence of ambiguity}

Every centroid estimator must assign a single reconstructed position to each observed cluster, even when multiple true interaction phases are consistent with the same observables. If the estimator resolves these ambiguities systematically rather than randomly, the preferred reconstructed position becomes dependent on the true subpixel phase. Consequently, the reconstructed phase density becomes non-uniform despite uniform illumination, producing subpixel phase bias.

This introduces an intrinsic trade-off between geometric specificity and translational symmetry. Estimators that exploit increasingly detailed cluster morphology or timing information can extract additional directional information from the cluster, but they also become increasingly sensitive to the discrete detector lattice and to the phase-dependent probabilities of different cluster morphologies. Simpler first-moment estimators generally preserve translational symmetry more naturally.

A further consequence is that exact entry-point reconstruction from detector observables is impossible in the general case. Even with a perfect forward model, distinct interaction points may produce overlapping detector responses, so deterministic centroid estimators must select a single reconstruction from an intrinsically non-unique inverse problem. Such deterministic resolution can introduce phase bias which becomes visible after rebinning if the reconstruction mapping does not preserve the appropriate intra-pixel symmetries.

\section{Implications for imaging fidelity}
\label{sec:3}

\subsection{Phase-dependent reconstruction bias}

A conventional ToT-weighted centroid approximates the first spatial moment of the deposited charge distribution and therefore varies smoothly with the measured observables~\cite{Khalil, Christodoulou_2024}. Consequently, it approximately preserves the reflection symmetry within the pixel and supports statistically uniform rebinning into a $2\times2$ virtual pixel grid, even without providing uniform intra-pixel phase density.

Within the framework developed here, centroid estimators that rely on discrete event-level decisions are expected to be more susceptible to phase-dependent reconstruction bias. Examples include selecting a particular hit pixel according to an extremal observable, topology classification, principal-axis fitting or extrapolation along a fitted axis, and machine-learning regressors trained solely to minimise localisation error.

Such methods exploit detailed event-level cluster structure, including directionality, elongation, and timing information that are not captured by a first-moment estimator~\cite{VANSCHAYCK2020113091}. This information, however, is encoded in a finite set of cluster observables, and so a finite set of discrete cluster morphologies. The probability of observing each morphology varies with the true subpixel phase, while the observed morphology of an individual event is discrete. The general mapping in eq.~\ref{E2.4} may therefore take the morphology-dependent form $\rhat=f_k(Z)$, with different reconstruction rules $f_k$ associated with different morphology classes $k$. As the relative probabilities of these classes vary across the pixel, so does the mixture of reconstruction rules applied, which can preferentially populate particular reconstructed subpixel phases. In a rebinned flat-field image, this phase-dependent occupancy can become visible as patterned structure such as a checkerboard.

This also explains why localisation accuracy alone does not predict imaging performance. Localisation metrics depend only on the marginal distribution of the residual $|\rhat-\rtrue|$, whereas checkerboarding and modulation transfer are governed by the spatial structure of the conditional reconstruction kernel $p(\rhat\mid\rtrue)$. Consequently, two centroid estimators may exhibit similar residual distributions while producing markedly different subpixel phase distributions and therefore different MTFs~\cite{VANSCHAYCK2020113091}.

\subsection{LSI systems and the meaning of MTF}

A general linear imaging system may be written as
\begin{align}
I_{\mathrm{out}}(\mathbf r)
=
\int
K(\mathbf r,\mathbf r')
I_{\mathrm{in}}(\mathbf r')
\,d\mathbf r',
\label{E3.2}
\end{align}
where $K(\mathbf r,\mathbf r')$ is the imaging kernel describing the response at the output position $\mathbf r$ to an input at $\mathbf r'$. A linear shift-invariant (LSI) system is the special case in which the kernel depends only on the relative displacement~\cite{goodman2005introduction},
\begin{align*}
K(\mathbf r,\mathbf r') = \mathrm{PSF}(\mathbf r-\mathbf r').
\end{align*}

Only in this case does the imaging operation reduce to a convolution, allowing the convolution theorem to be applied to eq.~\ref{E3.2} as $\widetilde I_{\mathrm{out}}(f)
=
\widetilde{\mathrm{PSF}}(f)\,
\widetilde I_{\mathrm{in}}(f).$ The optical transfer function is therefore
\begin{align*}
\mathrm{OTF}(\mathbf f) = \mathcal F\{\mathrm{PSF}\},
\end{align*}
and its magnitude defines the modulation transfer function,
\begin{align*}
\mathrm{MTF}(f) = |\mathrm{OTF}(f)|.
\end{align*}

For pixelated detectors, the system remains approximately shift invariant from pixel to pixel, but the centroid reconstruction may still depend on the true intra-pixel phase. The measured MTF can therefore be interpreted as an effective average over subpixel phases,
\begin{align}
\mathrm{MTF}_{\mathrm{eff}}(f)
\approx
\left<
\mathrm{MTF}(f\mid\phi_x,\phi_y)
\right>_{\phi_x,\phi_y}.
\end{align}

This approximation remains meaningful provided the reconstruction is approximately shift invariant at the pixel scale. When rebinning reveals explicit phase-dependent reconstruction bias, however, the reconstruction can no longer be described by a single translation-invariant PSF. Consequently, the convolution theorem no longer applies globally, and the rebinned image contains estimator-generated lattice frequencies in addition to transferred object frequencies. The rebinned MTF should therefore not be interpreted as an intrinsic measure of detector imaging performance.

\section{Experimental observations}

The proposed framework was evaluated using simulated 200~keV and 300~keV electron data together with measured 200~keV electron data. Four centroid estimators were compared in terms of localisation accuracy, modulation transfer function (MTF), and rebinnability: a ToT-weighted centroid, a maxToA centroid, a morphology-dependent 2D centroid, and a 3D track-extrapolation centroid.

The ToT-weighted centroid reconstructs the interaction position from the first spatial moment of the collected charge. The maxToA method assigns the interaction to the centre of the pixel with the latest time of arrival, roughly correlated to the earliest deposited charge in the sensor. The 2D and 3D methods additionally exploit cluster morphology, with the latter also incorporating an interaction-depth estimate to extrapolate the electron trajectory back to the detector surface.

\subsection{Localisation versus imaging performance}

The centroid estimators were first developed and tuned using simulated 200~keV and 300~keV electron data generated with Allpix$^2$. At this stage, neither ToT calibration nor timewalk correction was applied, allowing the intrinsic localisation performance of each estimator to be compared directly against the simulated interaction position.

\begin{figure}[htbp]
\centering
\includegraphics[width=0.7\linewidth]{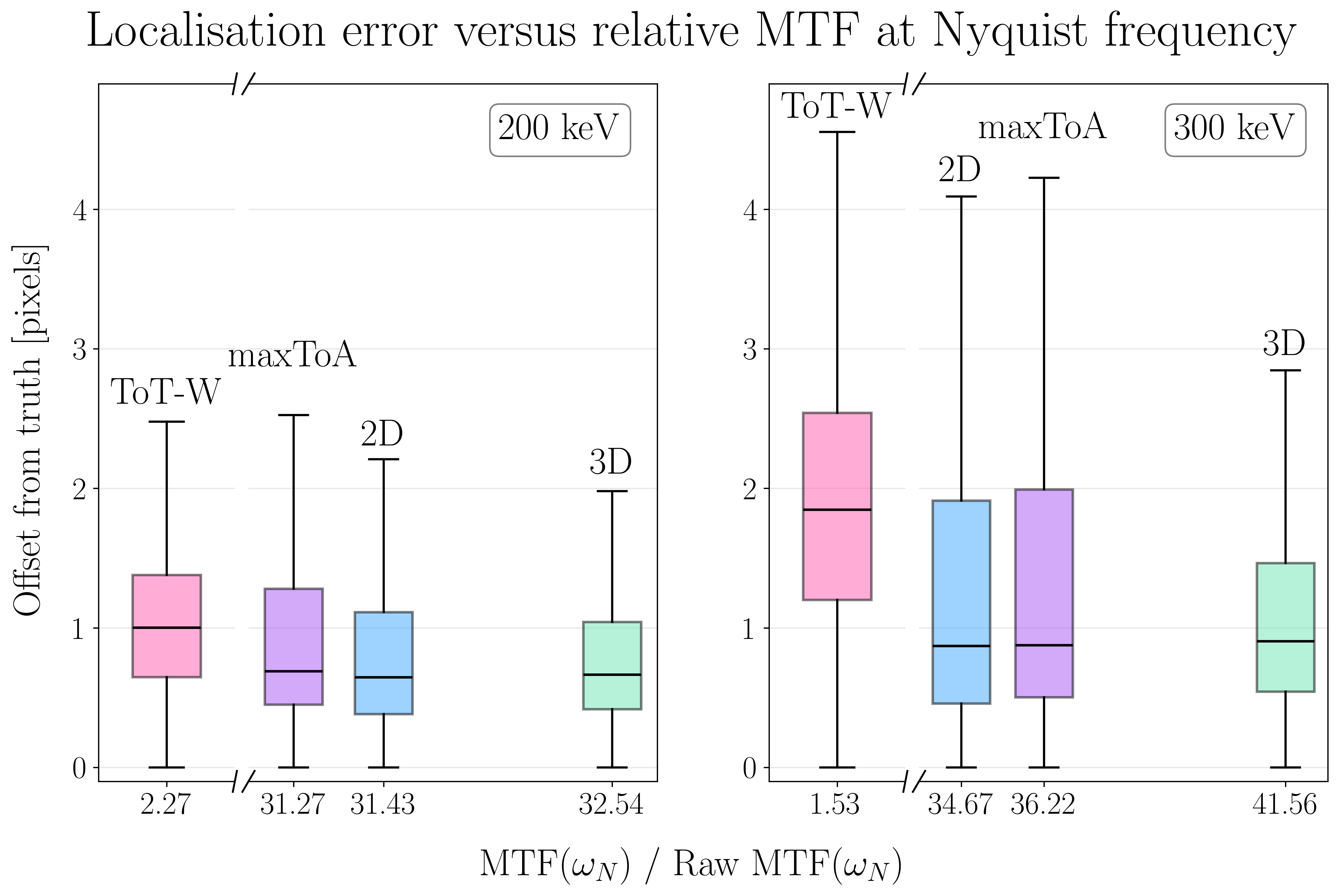}
\caption{Comparison of localisation and imaging performance for simulated 200~keV and 300~keV electrons. The box plots show the localisation error for each centroiding method. Boxes indicate the interquartile range, bold horizontal lines denote the median, and whiskers extend to the most extreme values within $1.5$ times the interquartile range. The MTF at Nyquist for each centroiding method is shown relative to the raw-data value. Statistical uncertainties on the MTF values are smaller than the plotted marker widths.\label{fig:localisation}}
\end{figure}

Figure~\ref{fig:localisation} compares the localisation error and relative MTF at Nyquist for the four centroid estimators at 200~keV and 300~keV. The relative MTF is defined as the ratio of the MTF values obtained using a centroiding method to those for the raw detector response. Localisation performance is represented by box plots of the distance between the reconstructed centroid and the true interaction point, positioned according to the corresponding relative MTF of each method. The morphology-dependent 2D and 3D centroid estimators reduce the localisation error compared with the conventional ToT-weighted centroid, with the 3D method achieving the smallest median error at both electron energies.

Although improved localisation generally coincides with larger relative MTF, the relationship is not monotonic. In particular, for 300~keV electrons the 2D centroid exhibits localisation performance comparable to the maxToA centroid while producing a lower MTF, a behaviour also reported for Timepix3 detectors~\cite{VANSCHAYCK2020113091}. Similarly, the improvement in MTF between the 2D and 3D methods is considerably smaller than the corresponding reduction in localisation error.

These observations demonstrate that localisation accuracy alone is insufficient to predict imaging performance, motivating the investigation of subpixel phase behaviour presented  in the following section.

\subsection{Real 200~keV imaging and rebinnability}

\begin{figure}[htbp]
\centering
\includegraphics[width=.7\textwidth]{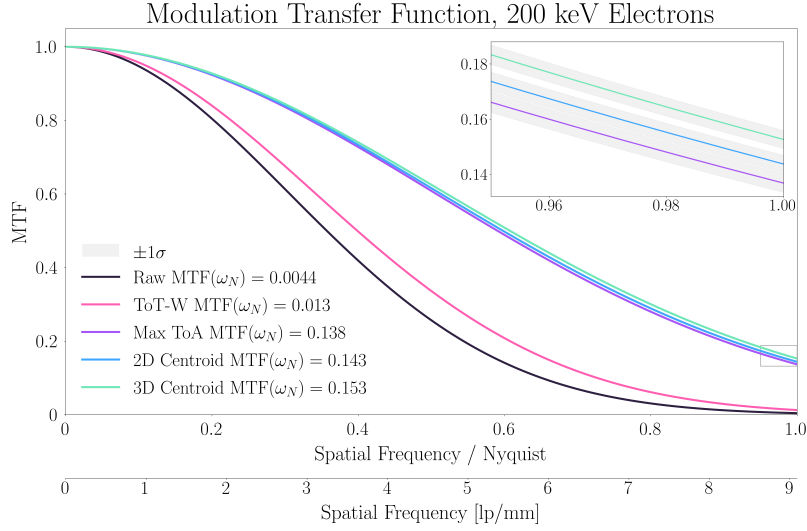}
\caption{Measured modulation transfer functions for 200~keV electrons reconstructed using the four centroiding strategies. The inset shows the MTF at Nyquist, highlighting the improvement obtained with the morphology-dependent centroid estimators over the maxToA centroid. The shaded bands indicate the $\pm1\sigma$ uncertainty.\label{fig:real_data}}
\end{figure}

The centroid estimators were finally applied to measured 200~keV electron data acquired using a Timepix4 detector bump-bonded to a $300~\upmu$m silicon sensor and operated in a JEOL CryoARM Z300FSC transmission electron microscope. The detector was read out using the MerlinT4 system. Figure~\ref{fig:real_data} compares the measured modulation transfer functions obtained using the four centroiding strategies. Beyond the ToT-weighted centroid, the maxToA, 2D, and 3D centroid estimators each substantially improves the detector MTF relative to the raw pixel response, demonstrating the practical benefit of centroid reconstruction for direct electron imaging.

Since centroiding improves the detector MTF, a natural next step is to investigate whether the reconstructed image can be further improved by rebinning the centroid positions into a finer virtual pixel grid. Such rebinning requires the reconstructed subpixel coordinates to preserve the approximate translational symmetry of the detector response, resulting in an approximately uniform population of the pixel cell and therefore statistically uniform virtual pixels.

\begin{figure}[htbp]
\centering
\begin{subfigure}{0.35\textwidth}
    \centering
    \includegraphics[width=\linewidth]{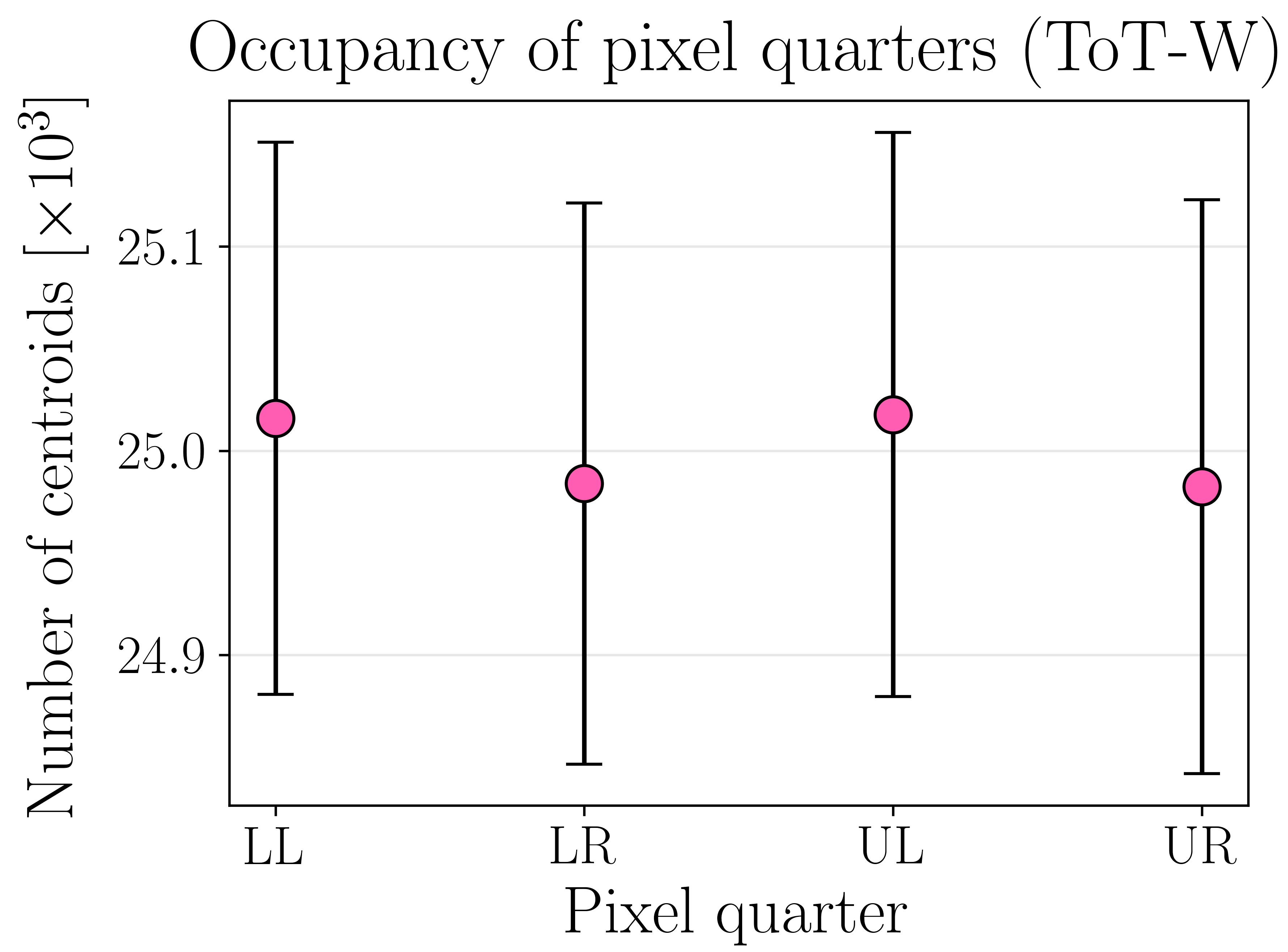}
    \caption{}
    \label{fig:intra:a}
\end{subfigure}
\hspace{0.1\textwidth}
\begin{subfigure}{0.35\textwidth}
    \centering
    \includegraphics[width=\linewidth]{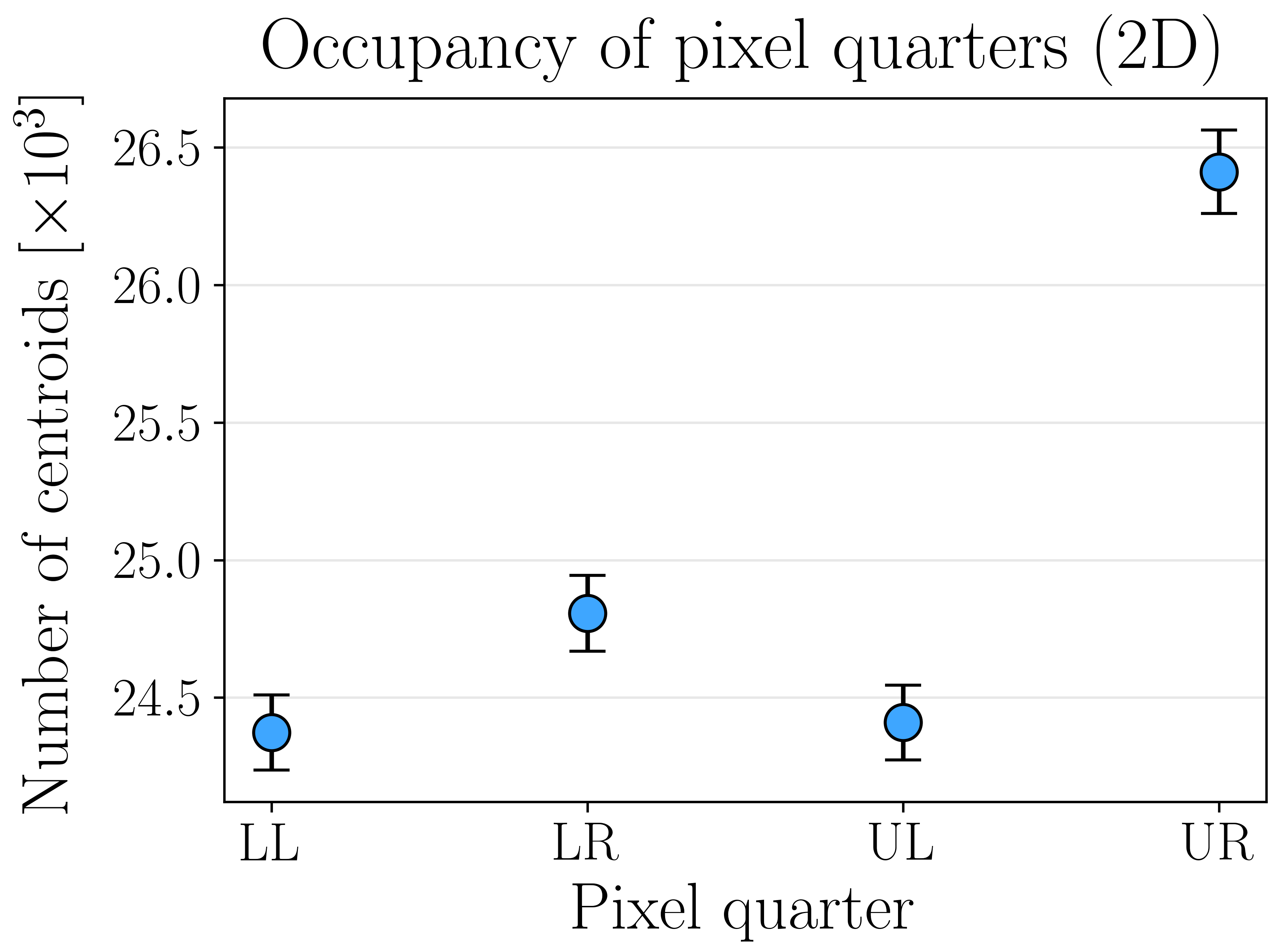}
    \caption{}
    \label{fig:intra:b}
\end{subfigure}
\vfill
\begin{subfigure}{0.3\textwidth}
    \centering
    \includegraphics[width=\linewidth]{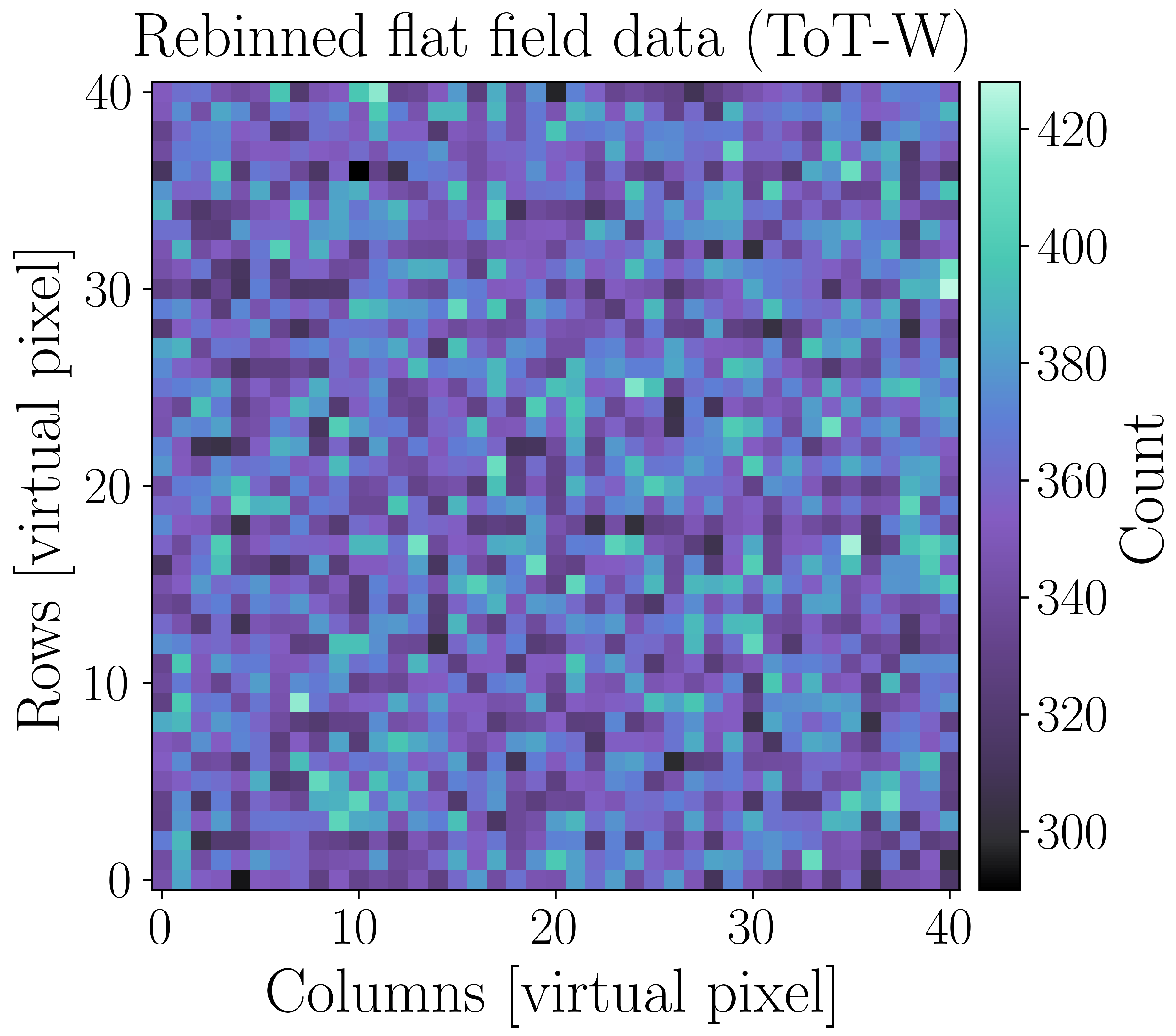}
    \caption{}
    \label{fig:intra:c}
\end{subfigure}
\hspace{0.15\textwidth}
\begin{subfigure}{0.3\textwidth}
    \centering
    \includegraphics[width=\linewidth]{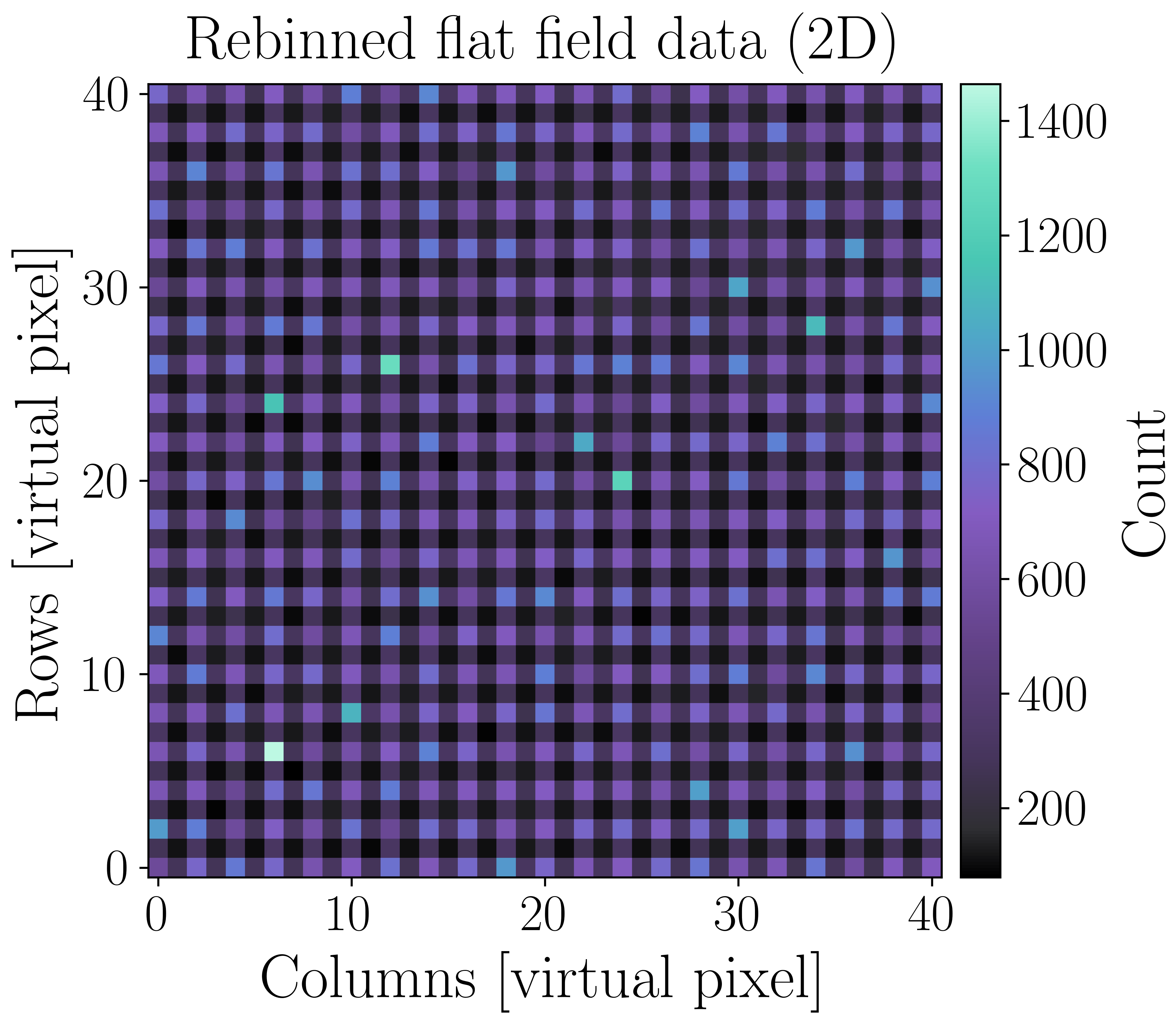}
    \caption{}
    \label{fig:intra:d}
\end{subfigure}
\caption{Comparison of intra-pixel symmetry and rebinnability for the ToT-weighted and morphology-dependent 2D centroid estimators. Mean occupancy of the four pixel quadrants for the (a) ToT-weighted centroid and (b) 2D centroid. Points show the mean quadrant occupancy across groups of $10^5$ reconstructed centroids, with error bars indicating the $\pm1\sigma$ standard deviation between groups. Flat-field data reconstructed with the (c) ToT-weighted centroid and the (d) 2D centroid, rebinned onto a $2\times2$ virtual pixel grid. \label{fig:intra}}
\end{figure}

To test this, the reconstructed centroid positions were analysed within the pixel cell. Figure~\ref{fig:intra:a} shows the mean occupancy of the four pixel quadrants obtained using ToT-weighted centroiding. The approximately equal quadrant occupancies do not imply that the reconstructed phase density is uniform within each quadrant. Rather, they show that the integrated probability assigned to each quadrant is approximately equal. This weaker condition is sufficient for the $2\times2$ rebinning considered here, producing statistically uniform virtual-pixel occupancies as shown in Figure~\ref{fig:intra:c}.

The corresponding quadrant occupancies for the morphology-dependent 2D centroid are shown in Figure~\ref{fig:intra:b}. Here, the reconstructed centroids occupy preferred subpixel locations, producing a distribution that is no longer approximately reflection symmetric. Consequently, the rebinned flat field develops the checkerboard artefact shown in Figure~\ref{fig:intra:d}.

These observations provide direct experimental evidence for the phase-dependent reconstruction bias introduced in Section~\ref{sec:3}. The limitation on imaging performance is therefore not solely the average distance between the reconstructed centroid and the true interaction point, but also the spatial organisation of the reconstructed centroids within the pixel cell. Once the reconstruction introduces periodic lattice structure, the rebinned image no longer represents the response of an approximately shift-invariant imaging system, and its measured MTF contains both transferred object frequencies and estimator-generated lattice frequencies.

Faithful spatial-frequency transfer requires preserving approximate intra-pixel translational symmetry to avoid phase-dependent reconstruction bias, maintain statistically uniform rebinnability, and ensure that the measured MTF remains an intrinsic measure of detector imaging performance.

\section{Conclusions and outlook}

This work proposes a unified interpretation of the relationship between localisation accuracy, modulation transfer function, and rebinnability for subpixel centroid reconstruction. We argue that observable-based centroid estimators solve an intrinsically non-unique inverse problem, and so can introduce reconstruction bias that depends on the electron entry phase. Consequently, localisation metrics alone are insufficient to predict detector imaging performance. Instead, the $2\times2$ rebinning considered here requires approximately equal integrated occupancies of the four pixel quadrants. A uniform intra-pixel phase density is a stronger condition that would be required for arbitrary finer rebinning.

The proposed framework provides a common interpretation for several observations reported in the literature. Convolutional neural-network centroiding for Timepix3 substantially improved localisation accuracy, while still requiring empirical sub-pixel redistribution to recover a uniform flat-field response, consistent with the presence of residual phase-dependent reconstruction bias~\cite{VANSCHAYCK2020113091}. This demonstrates that machine-learning estimators are not inherently immune to phase bias. When trained solely to minimise localisation error, they learn the statistical detector response rather than the underlying continuous interaction geometry. Additionally, the study illustrates that improved localisation does not strictly correlate with improved imaging performance, reinforcing the distinction between scalar localisation metrics and the spatial organisation of reconstructed subpixel coordinates.

Similar empirical correction methods based on the ToT-weighted centre of mass have successfully restored an approximately uniform subpixel response in Timepix3 detectors using pre-computed correction maps~\cite{Christodoulou_2024}. This approach is effective for diffusion-dominated events, where the residual subpixel bias is itself approximately symmetric and can therefore be compensated successfully by empirical redistribution. In contrast, the events considered here are dominated by electron scattering, making simple redistribution insufficient.

The magnitude of these phase-bias effects is expected to depend strongly on the information content of the measured cluster. The MÖNCH detector employs a 25~$\upmu$m pixel pitch, substantially smaller than that of Timepix4, producing considerably larger event clusters for the same electron energy. This allows deep-learning models to exploit richer spatial information and achieve much smoother localisation~\cite{Xie_2024}.

Future work with the Timepix4 will therefore focus on developing symmetry-preserving centroid estimators that explicitly incorporate translational symmetry into the reconstruction, rather than optimising localisation accuracy alone. More generally, the framework presented here suggests that future centroid design should be evaluated using both localisation and imaging metrics, together with quantitative measures of subpixel phase bias and rebinnability.


\bibliographystyle{JHEP}
\bibliography{biblio.bib}


\end{document}